\documentclass[showpacs,preprintnumbers,amsmath,amssymb,APSl,prd,nofootinbib,superscriptaddress, twocolumn]{revtex4-2}

\usepackage{amsmath}
\usepackage{amssymb}
\usepackage{amsfonts}
\usepackage{graphicx,bm}
\usepackage{dcolumn}
\usepackage[colorlinks=true]{hyperref}
\usepackage{epsf}
\usepackage{xfrac}
\usepackage{enumerate}
\usepackage{hhline}
\usepackage{array}
\usepackage{tabularx}
\usepackage{color}
\usepackage{float}
\usepackage[caption=false]{subfig}

\newcommand{\be}{\begin{equation}}
\newcommand{\ee}{\end{equation}}
\newcommand{\bea}{\begin{eqnarray}}
\newcommand{\eea}{\end{eqnarray}}
\newcommand{\beaa}{\begin{eqnarray*}}
\newcommand{\eeaa}{\end{eqnarray*}}

\def\be{\begin{equation}}
\def\ee{\end{equation}}
\def\bea{\begin{eqnarray}}
\def\eea{\end{eqnarray}}

\begin{document}
\title{Quasi-normal modes and echoes of generalized black hole bounces and their correspondence with shadows}

\author{Albert Duran-Cabac\'es} \email{albert.duran22@uva.es}
\affiliation{Department of Theoretical Physics, Atomic and Optics, and Laboratory for Disruptive Interdisciplinary Science (LaDIS), Campus Miguel Delibes, \\ University of Valladolid UVA, Paseo Bel\'en, 7,
47011 - Valladolid, Spain}

\author{Diego Rubiera-Garcia} \email{drubiera@ucm.es}
\affiliation{Departamento de F\'isica Te\'orica and IPARCOS,
	Universidad Complutense de Madrid, E-28040 Madrid, Spain}
    
\author{Diego S\'aez-Chill\'on G\'omez}
\email{diego.saez@uva.es} 
\affiliation{Department of Theoretical Physics, Atomic and Optics, and Laboratory for Disruptive Interdisciplinary Science (LaDIS), Campus Miguel Delibes, \\ University of Valladolid UVA, Paseo Bel\'en, 7,
47011 - Valladolid, Spain}
\affiliation{Departamento de F\'isica, Universidade Federal do Cear\'a (UFC), Campus do Pici, Fortaleza - CE, C.P. 6030, 60455-760 - Brazil}

\begin{abstract}

We study the quasi-normal modes (QNMs) of a family of generalized black bounces interpolating between regular black holes and traversable wormhole solutions according to a single extra parameter $a$. Firstly, working with a generic spherically symmetric space-time with arbitrary radial function and an anisotropic fluid matter source, the general equations for the gravitational waves are obtained. Then, we focus on such particular space-time metric and use the time-domain method to find the evolution of the QNMs with respect to the parameter $a$, finding larger frequencies and damped modes as $a$ grows. Furthermore we find that, for a gap in the values of $a$ for which no horizon is present but several photon spheres are, echoes are produced. Such echoes, which come from trapped modes in the potential well that are eventually leaked off for higher frequencies, appear as repetitions of the original wave but with modulated amplitude and decreased frequencies, and study their evolution with $a$. In addition, at the light of the correspondence recently discussed in the literature between QNMs and black hole imaging, we discuss the relation of the features of such echoes with those features (photon rings and shadows) of optical images from thin accretion disks. Despite working with simplified models and settings, our analysis provides useful insights on the usefulness of the correspondence for both gravitational waves and shadows.

\end{abstract}
%
%
\maketitle
%
%
%
\section{Introduction}

In the last few years the awakening of multi-messenger astronomy, namely, astronomy with different carriers (neutrinos, cosmic rays, light and gravitational waves) has paved the way for a golden era in gravitational physics \cite{Addazi:2021xuf}. Among the boost of research in the area propelled by this new possibility, the sub-sets of quantum gravity phenomenology \cite{AlvesBatista:2023wqm} and that of bridging ultra-compact objects to observational test are particularly appealing \cite{Cardoso:2019rvt}.

Ultra-compact objects can be defined as being compact enough to hold a photon sphere, namely, a unstable surface of bound null orbits. This includes, but it is not limited to, black holes (see the classification of \cite{Cardoso:2019rvt}). Concerning black holes and following the uniqueness theorems, the only axi-symmetric, electro-vacuum solution of Einstein's General Relativity (GR) equations corresponds to the Kerr-Newman, solely characterized by mass, angular momentum (the Kerr solution \cite{Kerr:1963ud}) and (though negligible in astrophysically realistic settings) electric charge. Such a solution is extremely successful for an observational point of view, being capable to explain the motion of S-stars around the galactic center \cite{Becerra-Vergara:2021gmx}, the data from X-ray spectroscopy off accretion disks \cite{Bambi:2016sac}, the inspiral, merger and ringdown phases of dozens of binary coallescences reported by LIGO-Virgo observatories \cite{Carullo:2018sfu,Isi:2019aib,LIGOScientific:2020tif}, and the imaging of the accretion flow around the M87 and Milky Way galaxies by the Event Horizon Collaboration \cite{EventHorizonTelescope:2021dqv,EventHorizonTelescope:2022xqj}. 


However, deep down in their innards, black holes carry a disturbing burden: the development of a space-time singularity, where some geodesics paths end and causal predictability is threatened, is unavoidable \cite{Senovilla:2014gza}. Attempts to resolve the central singularity of black holes have been proposed for decades appealing to different schemes, see e.g. \cite{Ayon-Beato:1998hmi,Carballo-Rubio:2019fnb,Torres:2022twv,Lan:2023cvz,Bueno:2024dgm,Alencar:2025jvl}. In recent times, and under the light of current and future observational opportunities, another question has been raised in the literature regarding such proposals: is it possible to bring to observational test any of such proposals in the same way that Kerr black hole does? \cite{Eichhorn:2021iwq}. To answer such a question represents multiple challenges on different fronts, in particular depending on the messenger one chooses to carry out such observational tests. In this sense, the intriguing correspondence reported through several works between quasi-normal modes of gravitational waves and photon rings and shadows of black hole imaging (in the geometrical optics approximation, see \cite{Stefanov:2010xz,Jusufi:2019ltj,Pedrotti:2024znu}), further empowered via greybody factors of scattering problems 
\cite{Konoplya:2024lir,Konoplya:2024vuj}, offers promising observational avenues to study this problem.

Quasi-normal modes (QNMs) are the vibrational states of black holes when subjected to a perturbation. Energy is radiated away under the form of gravitational waves while the presence of a horizon makes the system to be dissipative so a normal mode problem cannot be established. Accordingly, vibrational QNM frequencies of black holes are complex, with its real part describing its amplitude/frequency and the imaginary part its damping/instability time. They therefore provide useful insights on the features and dynamical stability of the corresponding black holes, particularly in the ringdown phase of binary mergers when setting into the final configuration state and allows for tests of the no-hair conjecture. Furthermore, within the correspondence pointed above, QNM frequencies real and imaginary parts are associated (in the geometrical optics approximation) with the two most salient features present in simulated black hole images, the shadow and the photon rings \cite{Bisnovatyi-Kogan:2022ujt,Staelens:2023jgr}, respectively. The latter is associated to surfaces of unstable boound geodesics (the photon shell or, in the spherically symmetric case, the photon sphere), which necessarily accompanies every black hole \cite{Carballo-Rubio:2024uas}. This opens the exciting possibility, should technology progress far enough, to perform cross-tests of the same object with two different messengers.

On the other hand, for horizonless, ultra-compact (in the sense of possessing surfaces of unstable, bound geodesics), the replacement of the black hole horizon by some sort of internal structure may be capable of producing a potential well. This is so because for objects without horizons photon spheres always come in pairs (see however \cite{Murk:2024nod}): an unstable and a stable (usually dubbed as anti-photon sphere) one. Whenever this is the case, such a structure would act as a sort of mirror, trapping low-frequency modes and gradually leaking high-frequency ones, leading to a series of repetitions of the main gravitational burst, or {\it echoes}. Given the fact that such echoes carry valuable information about the innards of the central region \cite{Cardoso:2016oxy}, they can potentially act as observational discriminators with respect to black holes \cite{Cardoso:2017cqb}. Furthermore, such echoes can be associated, on the black hole imaging side, with the creation of additional photon rings associated to light trajectories bouncing in the potential well. It should be pointed out, however, that instabilities may arise as the energy density increases due to the presence of such trapped modes within the potential well \cite{Cunha:2022gde}, though this is not always the case \cite{Marks:2025jpt}.

The main aim of this paper lies in studying the QNMs within a family of models which are based on the implementation of the black bounce proposal of Simpson and Visser \cite{Simpson:2018tsi}. The latter modifies canonical black hole solutions by promoting the radial coordinate to a radial bouncing function, which prevents the focusing of geodesics. Extensions of such a proposal have ranged far and wide, see e.g. \cite{Lobo:2020ffi,Mazza:2021rgq}. For the sake of this work we are interested in the proposal \cite{Lobo:2020ffi}. The new metric proposed there depends on a single extra parameter (the same as the original black bounce model) and provides a smooth transition between regular black hole solutions and traversable wormholes for a critical value of such a parameter. This allows us to study usual QNMs of black holes and echoes of horizonless compact objects in the same setting. Although the analysis of the QNM spectrum has previously been studied in the literature for the original Simpson-Visser proposal \cite{Franzin:2023slm}, the novelty of the metric pursued here is the presence of several peaks in the potential that, in the end, produce a characteristic kind of echoes. Then, the correspondence of such echoes with the optical appearance of the object is studied, where the emergence of additional photon rings is also associated with the peaks of the effective potential for null geodesics in the QNM-shadows correspondence.

The paper is organized as follows. In Sec. \ref{BBsection}, the metric for the black bounce spacetime is introduced. Sec. \ref{perturbations} is devoted to establish the general perturbations equations for a general spherically symmetric spacetime. Then, in Sec. \ref{QNMs}, the QNMs are computed for the particular black bounce spacetime considered here. In Sec.  \ref{ShadowsCorresp}, the correspondence with the optical counterpart is discussed. Finally, Sec. \ref{conclusions} gathers the conclusion of the paper.


\section{A Regular Black Bounce space-time}
\label{BBsection}

Our theoretical framework is based on the following line element
\begin{equation} \label{eq:lineel}
    ds^2=-A(r)dt^2+\frac{dr^2}{B(r)}+\Sigma^2(r)\left(d\theta^2+\sin \theta d\phi^2\right),
\end{equation}
with the functions defined in \cite{Lobo:2020ffi} as
\begin{equation} \label{eq:functions}
    A(r)=B(r)=1-\frac{2Mr^2}{(r^2+a^2)^{3/2}} \quad ; \quad \Sigma(r)=\sqrt{r^2+a^2}.
\end{equation}
where $M$ is the usual asymptotic (ADM) mass and a single extra parameter $a$ is included. The promotion of the radial coordinate $r$ to a non-zero radial function via the replacement $r \to \Sigma(r)$ was introduced by Simpson and Visser in \cite{Simpson:2018tsi}, rooted on the proposal for a wormhole solution of Ellis \cite{Ellis:1973yv}. The presence of the bounce prevents the focusing of geodesics, hence removing the presence of a singularity in terms of incompleteness of geodesics. Within GR, such a bounce can be generally sourced by a combination of non-linear electrodynamics and a (phantom) scalar field \cite{Bronnikov:2021uta}.

The geometry introduced above features several interesting properties. Horizons are simply obtained via the zeroes of $B(r)=0$, which is a sixth-degree equation that can be reduced to a third-degree equation in the coordinate $u=r^2$. Analyzing it one finds that for $a<a_{I}=4M/3\sqrt{3}$ there are two real solutions, $r_+>r_-$, corresponding to the event and inner horizons of a black hole. For $a=a_{I}$ such horizons merge into a single one, corresponding to an extreme black hole. And finally, if $a>a_\text{I}$ there are no real solutions, meaning that we are dealing with a horizonless compact object. The bounce appears at $r=0$, where the areal-radius of the two-spheres attains a minimum $\Sigma=a$. Therefore, solutions with $a<a_{I}$ can be interpreted as regular black holes and those with $a>a_{I}$ as traversable wormholes. 

Furthermore, traversable wormholes of the kind above are compact enough to hold photon spheres, provided that the parameter lies within the range $a_I<a_{II}$, where $a_{II}=2\sqrt{5}M/5$, as discussed in detail in Sec. \ref{ShadowsCorresp} below. This offers the possibility of studying both the echoes and the optical appearances of this family of objects.



\section{Perturbation equations}
\label{perturbations}

In this section we consider the general problem of perturbations in an arbitrary spherically symmetric background sourced by an anisotropic fluid, before going on in the next section to solve the corresponding QNM problem.

\subsection{General decomposition}

We first shall obtain the gravitational perturbation equations for a general spherically symmetric background of the form (\ref{eq:lineel}). To this end, we shall closely follow the derivation in \cite{Maggiore:2018sht}, which is the standard procedure used by Regge-Wheeler \cite{Regge:1957td} and Zerilli \cite{Zerilli:1970wzz}, with the corresponding amendments and additions that have been applied later in the literature, see e.g. \cite{Sago:2002fe}. The first step is to consider a small perturbation $h_{\mu \nu}$ around the background metric $\overline{g}_{\mu \nu}$, that is, $g_{\mu\nu}=\overline{g}_{\mu\nu}+h_{\mu\nu}$ with $\vert h_{\mu\nu} \vert \ll 1$. We assume that this perturbation is produced by a same-order perturbation of the energy-momentum tensor $\delta T_{\mu \nu}$, namely:
\begin{equation}
T_{\mu \nu}=\overline{T}_{\mu \nu}+\delta T_{\mu \nu}
\end{equation}
As a consequence, the Einstein tensor is perturbed in the following way:
\begin{align}
    G_{\mu\nu}=R_{\mu \nu}-\frac{1}{2} g_{\mu \nu} R=\overline{G}_{\mu \nu}+\frac{1}{2} \big( \nabla^\lambda \nabla_\mu h_{\lambda \nu} \nonumber \\
    + \nabla^\lambda \nabla_\nu h_{\lambda \mu}-\Box h_{\mu \nu}- \nabla_\nu \nabla_\mu h
    + \Box h \overline{g}_{\mu \nu} \\ - \nabla_\alpha \nabla_\beta h^{\alpha \beta} \overline{g}_{\mu \nu}-\overline{R} h_{\mu \nu}+\overline{g}_{\mu \nu} h^{\alpha \beta}\overline{R}_{\alpha \beta} \big), \nonumber
\end{align}
We next need to cast the perturbed Einstein field equations, $G_{\mu\nu}=8\pi G T_{\mu \nu} $. To this end we just use the background Einstein field equations, $\overline{G}_{\mu\nu}=8\pi G \overline{T}_{\mu \nu}$) to obtain the following equation for the gravitational perturbations:
\begin{align}
\nabla^\lambda \nabla_\mu h_{\lambda \nu} + \nabla^\lambda \nabla_\nu h_{\lambda \mu}-\Box h_{\mu \nu}- \nabla_\nu \nabla_\mu h + \Box h \overline{g}_{\mu \nu} \nonumber \\ - \nabla_\alpha \nabla_\beta h^{\alpha \beta} \overline{g}_{\mu \nu}-\overline{R} h_{\mu \nu}+\overline{g}_{\mu \nu} h^{\alpha \beta}\overline{R}_{\alpha \beta}=16 \pi \delta T_{\mu \nu}.
\label{eq_pertEFE}
\end{align}
Since the background metric is spherically symmetric, the axial (odd) and polar (even) perturbations will not couple each other. We label as axial perturbations those that under a parity transformation pick a factor $(-1)^{l}$, while polar perturbations are those that pick a factor $(-1)^{l+1}$ under the same parity transformation. Therefore, we can safely deal with each type of perturbation separately. However, for simplicity, we will only study the axial perturbations in this paper. Decomposing the gravitational perturbation equations into tensorial spherical harmonics and choosing de Regge-Wheeler gauge fixing, the axial perturbation reads (we employ a nomenclature similar to the one used in \cite{Maggiore:2018sht}):
\begin{widetext}
\begin{equation}
    h_{\mu v}^{\text {axial }}=\sum_{l, m}\left(\begin{array}{cccc}0 & 0 & \frac{h^{Bt}_{l m}(r, t)}{\sin \theta} \partial_\varphi  & -\sin \theta h^{Bt}_{l m}(r, t) \partial_\theta  \\ 0 & 0 & \frac{h^{B1}_{l m}(r, t)}{\sin \theta} \partial_\varphi  & -\sin \theta h^{B1}_{l m}(r, t) \partial_\theta  \\ \frac{h^{Bt}_{l m}(r, t)}{\sin \theta} \partial_\varphi   & \frac{h^{B1}_{l m}(r, t)}{\sin \theta} \partial_\varphi  & 0 & 0 \\-\sin \theta h^{Bt}_{l m}(r, t) \partial_\theta  & -\sin \theta h^{B1}_{l m}(r, t) \partial_\theta& 0 & 0\end{array}\right)Y_{l m}(\theta, \varphi).
\end{equation}
\end{widetext}
In a similar way, we can decompose the energy-momentum tensor into tensorial spherical harmonics (we do not assume, a priori, any type of matter perturbing the spacetime to keep the derivation as general as possible):
\begin{widetext}
\begin{equation}
    \delta T_{\mu v}^{\text {axial }}=\sum_{l, m}\left(\begin{array}{cccc}0 & 0 & \frac{s^{Bt}_{l m}(r, t)}{\sin \theta} \partial_\varphi  & -\sin \theta s^{Bt}_{l m}(r, t) \partial_\theta  \\ 0 & 0 & \frac{s^{B1}_{l m}(r, t)}{\sin \theta} \partial_\varphi  & -\sin \theta s^{B1}_{l m}(r, t) \partial_\theta  \\ \frac{s^{Bt}_{l m}(r, t)}{\sin \theta} \partial_\varphi   & \frac{s^{B1}_{l m}(r, t)}{\sin \theta} \partial_\varphi  & -\frac{1}{\sin \theta} s^{B2}_{l m}(r, t) X & \sin \theta s^{B2}_{l m}(r, t) W \\-\sin \theta s^{Bt}_{l m}(r, t) \partial_\theta  & -\sin \theta s^{B1}_{l m}(r, t) \partial_\theta& \sin \theta s^{B2}_{l m}(r, t) W & \sin\theta s^{B2}_{l m}(r, t) X\end{array}\right)Y_{l m}(\theta, \varphi),
    \label{eq_dTax}
\end{equation}
\end{widetext}
where the operators $X$ and $W$ are defined as
\begin{align}
    X&=2\partial_\theta \partial_\varphi - 2 \cot{\theta}\ \partial_\varphi, \\
    W&=\partial_\theta^2-\cot{\theta}\ \partial_\theta-\frac{1}{\sin^2\theta}\partial_\phi^2.
\end{align}
We point out that the tensorial spherical harmonics decomposition is useful to detach the angular dependence from the radial and temporal dependence. 

At this point, we shall use the \textsc{OGRe} package from \texttt{Mathematica} \cite{OGRe} to obtain the expression for the left-hand side of Eq.\eqref{eq_pertEFE}, while the right-hand side corresponds to Eq.\eqref{eq_dTax}. The axial part gives us three different equations. The first one is
\begin{align}
    \frac{B(r)\partial_r \left(h^{B1}_{lm}(r,t) A(r)\right)+A(r)\partial_r \left(h^{B1}_{lm}(r,t) B(r)\right)}{2}\nonumber \\
    -\partial_t h^{Bt}_{lm}(r,t)=-16 \pi G s^{B2}_{lm}(r,t) A(r),
    \label{eq_B2}
\end{align}
the second is
\begin{align}
    \left[\partial_r + 2\frac{\Sigma'(r)}{\Sigma(r)}-\frac{1}{2}\left(\frac{A'(r)}{A(r)}-\frac{B'(r)}{B(r)}\right) \right] \partial_t h^{B1}_{lm}(r,t)\nonumber \\
    -\Bigg[ \partial_r -\frac{1}{2}\left(\frac{A'(r)}{A(r)}-\frac{B'(r)}{B(r)}\right)\partial_r-2\frac{A'(r)\Sigma'(r)}{A(r)\Sigma(r)} \nonumber\\
    +\frac{\overline{R}}{B(r)}-\frac{l(l+1)}{B(r)\Sigma(r)^2}\Bigg]h^{Bt}_{lm}(r,t)=16\pi G \frac{s^{Bt}_{lm}(r,t)}{B(r)},
    \label{eq_Bt}
\end{align}
and the third is
\begin{align}
    \partial^2_t h^{B1}_{lm}(r,t)-\left( \partial_r-2\frac{\Sigma '(r)}{\Sigma(r)}\right) \partial_t h^{Bt}_{lm}(r,t) \nonumber \\
    -\Bigg(\frac{A(r) l(l+1)}{\Sigma(r)^2}-\overline{R} A(r)-
   A(r)B(r) \frac{\partial^2_r \left(\Sigma(r)^2 \right)}{\Sigma(r)^2}\nonumber \\
  -\partial_r \left(A(r) B(r) \right) \frac{\Sigma'(r)}{\Sigma(r)}\Bigg)=16\pi s^{B1}_{lm}(r,t)A(r),
  \label{eq_B1}
\end{align}
where $\overline{R}$ is the background Ricci curvature scalar, whose expression reads explicitly as 
\begin{eqnarray}
    \overline{R}&=&-\frac{B(r) A''(r)}{A(r)}-\frac{A'(r) B'(r)}{2 A(r)}\nonumber \\
    &-&\frac{2 B(r) A'(r) \Sigma '(r)}{A(r) \Sigma (r)}+\frac{B(r) A'(r)^2}{2 A(r)^2}-\frac{2 B'(r) \Sigma '(r)}{\Sigma (r)}\nonumber\\&-&\frac{4 B(r) \Sigma ''(r)}{\Sigma (r)}-\frac{2 B(r) \Sigma '(r)^2}{\Sigma (r)^2}+\frac{2}{\Sigma (r)^2}
\end{eqnarray}

Note that Eqs.\eqref{eq_Bt} and \eqref{eq_B1} are defined for $l\geq 1$, while Eq.\eqref{eq_B2} is only defined for $l \geq 2$. Moreover, Eqs.\eqref{eq_B2}, \eqref{eq_Bt} and \eqref{eq_B1} are not completely independent, since the conservation of the energy-momentum, $\nabla_\mu T^{\mu}_{\nu}=0$, relates the three of them. The perturbation of the energy-momentum tensor satisfies also the perturbed version of the conservation of the energy-momentum tensor:
\begin{equation}
\nabla_\mu \delta T^{\mu}_{\nu}+\delta\Gamma^{\mu}_{\mu \lambda}\overline{T}^\lambda_\nu-\delta\Gamma^{\lambda}_{\mu \nu}\overline{T}^\mu_\lambda=0
\end{equation}
Therefore, only two of the above equations are needed to completely describe the system. Inserting $\partial_t h^{Bt}_{lm}(r,t)$ from Eq.\eqref{eq_B2} into Eq.\eqref{eq_B1} we obtain:
\begin{align}
    \left(2\frac{\Sigma'(r)}{\Sigma(r)}A(r)B(r)-\frac{3}{2}\partial_r\left(A(r)B(r)\right)\right)\partial_r h^{B1}_{lm}(r,t) \nonumber \\
    +\Bigg(\frac{A(r)}{\Sigma(r)^2}l(l+1)-\overline{R}A(r)-\frac{A(r)B(r)}{\Sigma(r)^2}\partial_r^2\left(\Sigma(r)^2\right)\nonumber \\
    -\frac{1}{2}\partial^2_r \left(A(r)B(r)\right)\Bigg)h^{B1}_{lm}(r,t) +\partial_t^2 h^{B1}_{lm}(r,t)\nonumber \\-A(r)B(r)\partial_r^2h^{B1}_{lm}(r,t)=-\frac{\Sigma(r)}{\sqrt{A(r)B(r)}}S^{\text{ax}}_{lm}(r,t),
    \label{eq_preSchroAx}
\end{align}
where we have introduced the object
\begin{align}
    S^{\text{ax}}_{lm}(r,t)=-16 \pi G\frac{\sqrt{A(r)B(r)}}{\Sigma(r)}\Bigg[ s^{B1}_{lm}(r,t)A(r)\nonumber \\
    +\left(\partial_r-2\frac{\Sigma'(r)}{\Sigma(r)} \right) s^{B2}_{lm}(r,t)A(r)\Bigg]
\end{align}

At this point, it is convenient to define a new coordinate $r^*(r)$ and a Regge-Wheeler variable $Q_{lm}(r,t)$ according to the definitions:
\begin{align}
    \partial_{r^*}&=\sqrt{A(r)B(r)} \partial_r  \label{eq_tortoise}\\
    \quad Q_{lm}(r,t)&=\frac{\sqrt{A(r)B(r)}}{\Sigma(r)}h^{B1}_{lm}(r,t).
\end{align}
We point out that $r^*$, usually known as the tortoise coordinate, maps the region $r\in (r_h,+\infty)$ into $r^*\in (-\infty,+\infty)$, a transformation that will pay off later. With these variables, Eq.\eqref{eq_preSchroAx} can be written as:
\begin{equation}
    \left(\partial_{r^*}^2-\partial_t^2-V_l(r) \right) Q_{lm}(r,t)=S^{\text{ax}}_{lm}(r,t),
    \label{eq_RW0}
\end{equation}
with the corresponding potential:
\begin{eqnarray}
    V_l(r)&=&A(r)\Bigg[ \frac{l(l+1)}{\Sigma(r)^2}-\overline{R}\nonumber \\
    &-&\frac{3}{\Sigma(r)}\sqrt{\frac{B(r)}{A(r)}}\partial_r \left(\sqrt{A(r)B(r)}\Sigma'(r)\right)\Bigg].
\end{eqnarray}
Therefore, we have obtained a wave-type equation to properly understand the properties of the gravitational perturbation. In order to keep progressing, we perform a Fourier transform with respect to the time variable:
\begin{equation}
    Q_{lm}(r,t)=\frac{1}{2\pi}\int_{-\infty}^{\infty} d\omega\  \tilde{Q}_{lm}(r,\omega) e^{-i \omega t},
    \label{eq_fourier}
\end{equation}
which transforms the gravitational perturbation equation into a Schrödinger-like equation:
\begin{equation}
    \left(\partial_{r*}^2+\omega^2-V_l(r) \right) \tilde{Q}_{lm}(r,\omega)=\tilde{S}^{\text{ax}}_{lm}(r,\omega).
    \label{eq_RWf0}
\end{equation}
The above equation is only valid for $l \geq 2$, in agreement with the fact that the equations used above are only defined in this case. Should we want to make an analysis for $l<2$, we should re-do the procedure to obtain the expression for the potential.

\subsection{Anisotropic fluid}

Let us now assume that the general metric we are working with is generated by an anisotropic energy-momentum tensor, which is general enough a source to encompass many case of physical interest (including scalar and electromagnetic fields).  This would the case provided that $G_t^t\neq G_r^r\neq G_\theta^\theta=G_\varphi^\varphi$, while all the other components of the Einstein tensor are zero. This is so because the energy-momentum tensor of such a fluid can be written the following way
\begin{eqnarray}
    T_{\mu \nu}&=&(\rho(r)+p_{\perp}(r))u_\mu u_\nu\nonumber \\
    &+&(p_{\parallel}(r)-p_{\perp}(r))x_\mu x_\nu+p_{\perp}(r) g_{\mu \nu},
    \label{eq_anis}
\end{eqnarray}
where $\rho (r)$ is the energy density, $p_{\parallel}(r)$ is the longitudinal (radial) pressure and $p_{\perp}(r)$ is the transversal (angular) pressure of the fluid. Moreover, $u_\mu$ is a unitary time-like four-vector ($u_\mu u^\mu=-1$), while $x_\mu$ is a unitary space-like four-vector ($x_\mu x^\mu=1$) which has radial direction and is transverse to $u_\mu$ ($x_\mu u^\mu=0$). Using the normalization above, in the spherically symmetric line element (\ref{eq:lineel}) the components of these vectors can be written as 
\begin{equation}
    u^\mu=\left(\begin{array}{c} \sqrt{A(r)} \\ 0 \\  0 \\ 0 \end{array}\right), \quad x^\mu=\left(\begin{array}{c} 0 \\ \frac{1}{\sqrt{B(r)}} \\  0 \\ 0 \end{array}\right).
\end{equation}

If we perturb the background variables at first order in Eq.\eqref{eq_anis}, then we obtain the following expression for the energy-momentum perturbation:
\begin{eqnarray}
\delta T_{\mu \nu}&=&\left (\overline{\rho}(r)+\overline{p}_{\perp}(r) \right)\left (\delta u_{\mu}\overline{u}_{\nu}+\overline{u}_{\mu}\delta u_{\nu} \right)\nonumber \\
&+&\left (\overline{p}_{\parallel}(r)-\overline{p}_{\perp}(r) \right)\left (\delta x_{\mu}\overline{x}_{\nu}+\overline{x}_{\mu}\delta x_{\nu} \right)\nonumber \\
&+&\overline{p}_{\perp}(r) h_{\mu \nu}+\left( \overline{g}_{\mu \nu}+\overline{u}_{\mu}\overline{u}_{\nu}-\overline{x}_{\mu}\overline{x}_{\nu}\right)\delta p_{\perp}\nonumber \\
&+&\overline{u}_{\mu}\overline{u}_{\nu} \delta \rho +\overline{x}_{\mu}\overline{x}_{\nu}\delta p_{\parallel},
\label{eq_dT}
\end{eqnarray}
whose axial perturbations read:
\begin{eqnarray}
    \delta u^\text{ax}_\mu&=&\sum_{l, m}-\frac{1}{2}\sqrt{A(r)}\left(\begin{array}{c} 0 \\0 \\  \frac{1}{\sin \theta}u^{(3)}_{lm}(r,t)\partial_\varphi\\ -\sin \theta u^{(3)}_{lm}(r,t)\partial_\theta \end{array}\right) Y_{lm}(\theta,\varphi) \\
    \delta  x^\text{ax}_\mu &=&\sum_{l, m}\frac{1}{2\sqrt{B(r)}}\left(\begin{array}{c} 0 \\ 0 \\  \frac{1}{\sin \theta}x^{(3)}_{lm}(r,t)\partial_\varphi\\ -\sin \theta x^{(3)}_{lm}(r,t)\partial_\theta \end{array}\right) Y_{lm}(\theta,\varphi),
\end{eqnarray}
and:
\begin{align}
    \delta \rho&= \sum_{l, m} \delta \rho_{lm} (r,t) Y_{lm} (\theta, \varphi) \\
    \delta p_{\parallel}&= \sum_{l, m} {\delta p_{\parallel}}_{lm} (r,t) Y_{lm} (\theta, \varphi) \\
    \delta p_{\perp}&= \sum_{l, m} {\delta p_{\perp}}_{lm} (r,t) Y_{lm} (\theta, \varphi).
\end{align}
These perturbed quantities can next be expanded into scalar spherical harmonics and vectorial spherical harmonics (for the explicit expressions of such harmonics see \cite{Maggiore:2007ulw}), in order to separate the angular dependency of the free functions.

The background variables are spherically symmetric and can be written, in a comoving frame, as
\begin{equation}
   \overline{T}^{\mu}{}_\nu=\left(\begin{array}{cccc} -\overline{\rho}(r) & 0 & 0 & 0 \\ 0 & \overline{p}_{\parallel}(r) & 0 & 0 \\ 0 & 0 &\overline{p}_{\perp}(r) & 0 \\ 0 & 0 & 0 &\overline{p}_{\perp}(r)\end{array}\right),
\end{equation}
and are related, via the background Einstein equations, to the line element quantities as:
\begin{eqnarray}
    \overline{\rho}(r)&=&-\frac{B'(r) \Sigma '(r)}{8 \pi  G \Sigma (r)}-\frac{B(r) \Sigma ''(r)}{4 \pi  G \Sigma (r)}  -\frac{B(r) \Sigma '(r)^2}{8 \pi  G \Sigma (r)^2}\nonumber \\
  &+&\frac{1}{8 \pi  G \Sigma (r)^2},\\
   \overline{p}_{\parallel}(r)&=&\frac{B(r) A'(r) \Sigma '(r)}{8 \pi  G A(r) \Sigma (r)}+\frac{B(r) \Sigma '(r)^2}{8 \pi  G \Sigma (r)^2}-\frac{1}{8 \pi  G \Sigma (r)^2},\\
    \overline{p}_{\perp}(r)&=&\frac{B(r) A''(r)}{16 \pi  G A(r)}+\frac{A'(r) B'(r)}{32 \pi  G A(r)}+\frac{B(r) A'(r) \Sigma '(r)}{16 \pi  G A(r) \Sigma (r)}\nonumber \\
    &-&\frac{B(r) A'(r)^2}{32 \pi  G A(r)^2}+\frac{B'(r) \Sigma '(r)}{16 \pi  G \Sigma (r)}+\frac{B(r) \Sigma ''(r)}{8 \pi  G \Sigma (r)}.
\end{eqnarray}
These expressions define the corresponding equation of state for the matter sector, which is necessary to have a closed system of determined equations. Note also that the perturbed energy-momentum has a dependency on the metric perturbation. For this reason, it is necessary to rewrite Eq.\eqref{eq_RWf0} in order to explicitly set aside the contribution of the metric from the matter perturbation variables. Comparing Eqs.\eqref{eq_dTax} and \eqref{eq_dT}, and performing a Fourier transform in the former, we can obtain the following equivalences:
\begin{align}
    \tilde{s}^{Bt}_{lm}(r,\omega)&=\overline{p}_{\perp}(r) \tilde{h}^{Bt}_{lm}(r,\omega)+A(r)\frac{\overline{p}_{\perp}(r)+\overline{\rho}(r)}{2}\tilde{x}_3(r,\omega) ,\\
    \tilde{s}^{B1}_{lm}(r,\omega)&=\overline{p}_{\perp}(r) \tilde{h}^{B1}_{lm}(r,\omega)+\frac{\overline{p}_{\perp}(r)-\overline{p}_{\parallel}(r)}{2B(r)}\tilde{x}_3(r,\omega),\\
    \tilde{s}^{B2}_{lm}(r,\omega)&=0.
\end{align}
Therefore, Eq.\eqref{eq_RWf0} can be rewritten as
\begin{align}
    \left(\partial_{r*}^2+\omega^2-V^{RW}_l(r) \right) \tilde{Q}_{lm}(r,\omega)\nonumber\\
    =8 \pi G \sqrt{\frac{A(r)}{B(r)\Sigma(r)^2}} p_{\parallel}(r) \tilde{x}^{(3)}_{lm}(r,\omega),
\end{align}
where the potential term is now the Regge-Wheeler one for a general spherically symmetric spacetime with $l\geq 2$, that is
\begin{eqnarray} \label{eq:RWpot}
    V^{RW}_l(r)&=&A(r)\Bigg[ \frac{l(l+1)-2}{\Sigma(r)^2} \\
    &+&\Sigma(r)\sqrt{\frac{B(r)}{A(r)}}\partial_r \left(\sqrt{A(r)B(r)}\partial_r\left(\frac{1}{\Sigma(r)}\right)\right)\Bigg]. \nonumber
\end{eqnarray}
This result coincides with the one found e.g. in \cite{Feng:2024ygo, Franzin:2023slm, Konoplya:2024lch} In this sense, we point out that the expressions found in \cite{Franzin:2023slm, Konoplya:2024lch} correspond to a different coordinate system but they can be recovered with the appropriate change of coordinates from our own results.

\section{Quasi-normal Modes}
\label{QNMs}
Eq.\eqref{eq_RWf0} does not have stationary states. Instead, it has vibrational modes that decay in time, which are called quasi-normal modes (QNMs). Their frequencies $\omega=\omega_R +i \omega_I$ are therefore found to be complex upon solving of the equation. The real part $\omega_R$ depicts the oscillatory behavior, while the imaginary part $\omega_I$ describes the decay rate of the perturbation. The simplest way to compute them is considering a scalar test-particle field as a matter perturbation \cite{Franzin:2023slm}, since scalar perturbations couple to the polar sector, so the axial sector remains unperturbed,  $\tilde{x}^{(3)}_{lm}(r,\omega)=0$. Therefore, Eq.\eqref{eq_RWf0} has the form of a wave equation with a potential, that is
\begin{align}
   \left [\partial_{r*}^2-\left(\omega^2+V^{RW}_l(r) \right)\right]\phi(r,\omega)=0 .
    \label{eq_quasinormal}
\end{align}
The boundary conditions of the equation are a natural choice considering the physical nature of the system. In this sense, the QNMs are purely ingoing at the horizon ($r^*\rightarrow -\infty$) and purely outgoing at spatial infinity ($r^*\rightarrow \infty$). This means that QNMs do not reflect at the horizon while are radiated away at asymptotic infinity, thus leaving their domain at both boundaries. Mathematically speaking, this amounts to the conditions
\begin{equation}
    \phi(r^*\rightarrow-\infty,\omega)\sim e^{-i\omega r^*}, \quad \phi(r^*\rightarrow\infty,\omega)\sim e^{i\omega r^*} \ ,
    \label{eq_boundary}
\end{equation}
respectively. Note that, using the convention in Eq.\eqref{eq_fourier}, we expect the imaginary part to be negative if we want stability on the perturbed system. However, this might seem to be problematic, since it implies that the QNMs amplitudes blow up both at infinity and at the event horizon, as set by Eq.\eqref{eq_boundary}. However, as QNMs should be thought as quasi-stationary states, they are excited at a particular time and decay exponentially in time, avoiding the issue of their infinite amplitude at the mentioned boundaries \cite{Berti:2009kk}.

Several methodologies to compute the QNM either numerically or semi-analytically have been proposed (see reviews on methodologies in \cite{Konoplya:2011qq, Pani:2013pma,Franchini:2023eda}). Among the main procedures, direct integration and the time domain analysis are often used due to their simplicity. Direct integration is great to compute overtones, but it is not as good to obtain other dynamical information of the perturbations. The time domain methodology, instead, is great to perform a full analysis on the perturbation evolution. It consists of evolving an initial wavepacket through Eq.\eqref{eq_RW0} with $x^{(3)}_{lm}(r,t)=0$ and studying its temporal evolution at some fixed point $r=r_o$. The wavepacket vibration must be able to be broken down into its vibrational modes (i.e. the QNMs). At late times, after a transient state, the dominant mode should be the fundamental one, since it is the most long-lived (i.e., it has the smallest $\omega_I$). This way, by measuring the vibration frequency at late times, we obtain the fundamental QNM. If we want to compute overtones, we should use complementary methods once we have the fundamental mode.

\subsection{Numerical setup and time-domain analysis}

In this work, we use the time-domain method to obtain the QNMs, focusing primarily on the fundamental mode. This approach allows us to obtain more information about the dynamical evolution, such as echoes and late-time tails. Therefore, in order to solve Eq.\eqref{eq_RW0}, we first perform a change of variables $u=t-r^*$ and $v=t+r^*$. These are the so-called light-cone variables, and they are specifically useful to avoid reflections on the computational domain besides simplifying the wave equation, which is reduced to:
\begin{equation}
   \left( 4\frac{\partial^2}{\partial v \partial u}+V(u,v)\right)\phi (u,v)=0.
\end{equation}
Therefore, considering a squared grid for $u$ and $v$ with a step-size of $h$, the integration scheme reads \cite{Gundlach:1993tp, DuttaRoy:2019zvw}:
\begin{align}
\phi(u+h,v+h)=&\phi(u,v+h)+\phi(u+h,v)\nonumber \\&-\phi(u,v)-\frac{h^2}{8}V(u,v)\big[\phi(u,v+h)\nonumber \\&+\phi(u+h,v)\big].
\label{eq_numsteps}
\end{align}
Then, with initial conditions for the null surfaces $u=u_0$ and $v=v_0$, we can obtain the evolution of $\phi(u,v)$. The initial conditions should not affect the result of the QNMs in the ringdown stage \cite{Molina:2010fb,Chavda:2024awq}. Consequently, for easiness, we will use a Gaussian distribution for $\phi(u=0,v)$ of the form
\begin{equation}
    \phi(u=0,v)=e^{-\frac{(v-v_c)^2}{2\sigma^2}},
    \label{eq_initial}
\end{equation}
and a constant value for $\phi(u,v=0)=\phi(u=0,v=0)$.

Since the tortoise coordinate does not have in this case an analytical expression in terms of $r$, we need to compute it numerically in order to obtain $V(r^{*})$ and, subsequently, $V(u,v)$. Therefore, we need to guarantee some accuracy. That is not an easy task, given the fact that $r^*(r)$ has a singularity at the horizon. For this reason, we solve numerically the non-linear equation \eqref{eq_tortoise} for $r(r^*)$, as is done in e.g. \cite{Konoplya:2023ahd} and then we obtain $V(r(r^*))$.

\begin{figure}[t!]  \includegraphics[width=\columnwidth]{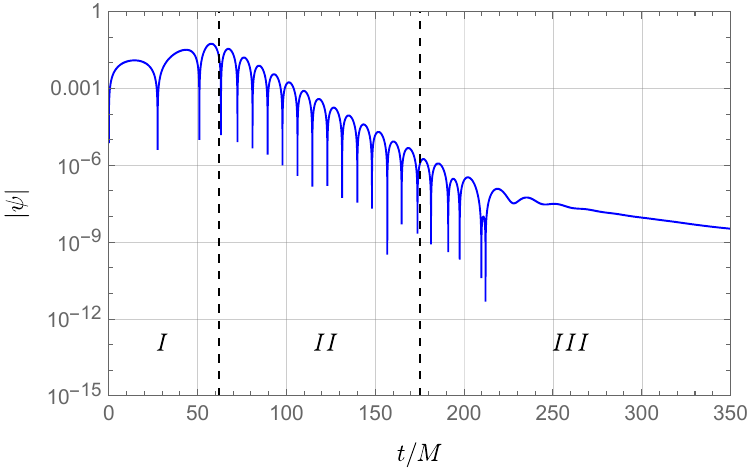}
\caption{Evolution of a Gaussian distribution centered in $v_c=50M$, observed at $r^*_o=150M$. We have used the Schwarzchild Regge-Wheeler potential (i.e. we set $a=0$). In order to observe the power-law tail behaviour, the observer was placed far enough from the perturbation, as in \cite{Kyutoku:2022gbr}.}
\label{fig_timeprofile}
\end{figure}

After computing the full evolution for $\phi(u,v)$, we can extract a time profile by undoing the change of coordinates and defining $\psi(t)=\phi(r^*=r^*_o,t)$. This time profile, whose modulus is represented in Fig. \ref{fig_timeprofile} will consist of three temporal phases. The first one is the transitory phase (region $I$), the behaviour of which is strongly related to the initial conditions. After this phase, the only QNM that survives is the fundamental one, dominating the whole region $II$. Finally, provided they are present, power-law tails may appear for observers far enough from the initial perturbation $III$ \cite{Nollert:1999ji,Leaver:1986gd}.

Having obtained the time profile, our next goal is to compute the QNM frequencies at late times. To this end we shall use the Prony fitting method and follow the methodology of \cite{Konoplya:2011qq,Berti:2007dg}. We assume that the time profile within this period can be described using the following decomposition of damped exponentials:
\begin{equation}
    \psi(t)=\sum_{n=0}^N C_n e^{-i\omega_n t}.
    \label{eq_damped}
\end{equation}

Therefore, we get $N$ values for the coefficients $\omega_n$ and $C_n$. 

At late times, the predominant frequency corresponds to the fundamental QNM. With an appropriate choice of the time region to be studied (region $II$ in Fig. \ref{fig_timeprofile}), the expansion in Eq.\eqref{eq_damped} is almost monochromatic, i.e., one of the frequencies $\omega_n$ has a significantly dominant amplitude $C_n$ over the other ones. Actually, if the studied time region is sufficiently monochromatic, the amplitudes of the $N-1$ non-fundamental modes fall below the numerical precision of our time profile, making it easier to identify the fundamental one.
\subsection{Results}

\begin{figure}[t!]  \includegraphics[width=0.9\columnwidth]{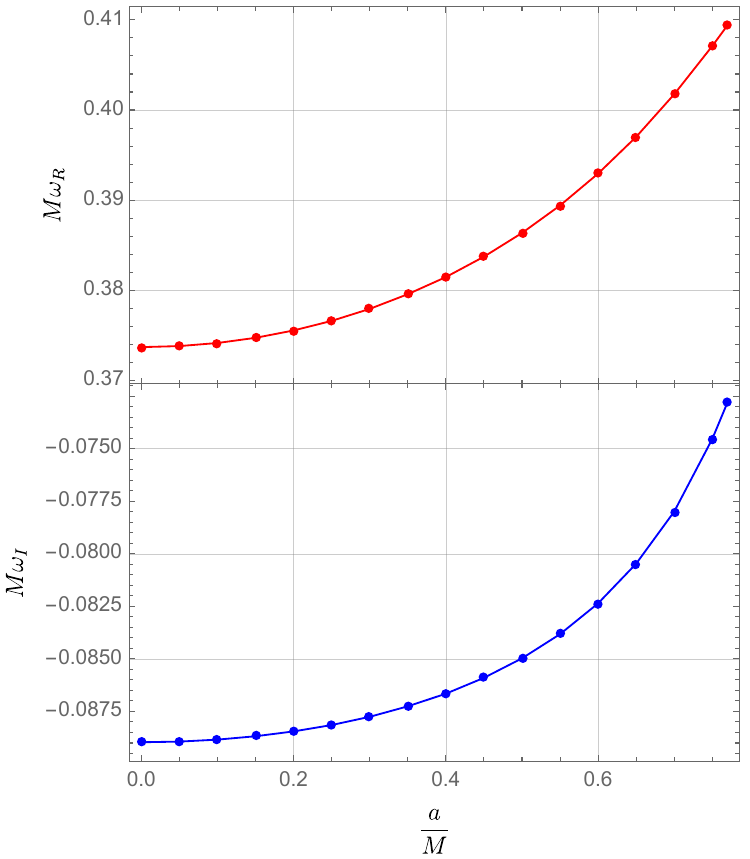}
\caption{Real part (top, red) and imaginary part (bottom, blue) of the QNM as a function of the parameter $a$ in the case where at least one horizon exists, $0 \leq a \leq a_{I}$. The last point corresponds to the extremal case ($a=a_\text{I}$).}
\label{fig_modes}
\end{figure}

As discussed in previous sections, the QNMs are computed using the time-domain method. We prepare an initial state as in Eq.\eqref{eq_initial} with $\sigma=0.25M$, $v_c=10M$, and we evolve it through Eq.\eqref{eq_numsteps}. In order to take advantage of the numerical precision, we use different values of the numerical domain of $u$ and $v$, in addition to different values for the numerical step $h$ depending on the behaviour of the QNM. The main reason for that choice is the fact that long-lasting modes (i.e. small $|\omega_I|$) need less numerical precision and more numerical range, while short-lasting (large $|\omega_I|$) modes behave the other way around.

For both the black hole case with two non-degenerate horizons and the extremal case (i.e. within the range $a\leq a_I)$, we have used $h=0.005$ and $u, v \in [0, 400\ M]$, and the results are shown in Fig. \ref{fig_modes}. We can see how the case in which $a=0$ agrees with the Schwarzschild fundamental QNM  found in previous works both semi-analytically and numerically \cite{Mamani:2022akq,Chandrasekhar:1975zza}. For growing values of $a$ we find that the real part of the QNM frequency steadily increases, while the absolute value of the imaginary part decreases. This behaviour is similar to the one reported for other models of regular black holes,  such as the Bardeen one \cite{Franzin:2023slm}. The main difference is that this increase (decrease) of the real (imaginary) parts is more gradual, accelerated when nearing the extreme black hole case, where significant differences are found.

\begin{figure}[t!]
  \includegraphics[width=0.9\columnwidth]{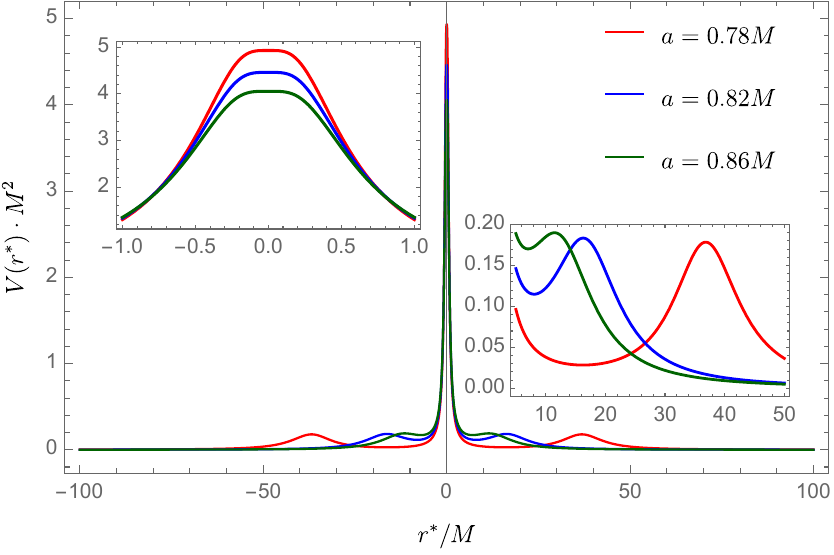}
\caption{Shape of the Regge-Wheeler potential for the axial perturbations in the space-time defined by the functions (\ref{eq:functions}) for three values of the parameter $a>a_\text{lim}$. }
\label{fig_peaks}
\end{figure}

\begin{figure}[t!]
  \includegraphics[width=0.9\columnwidth]{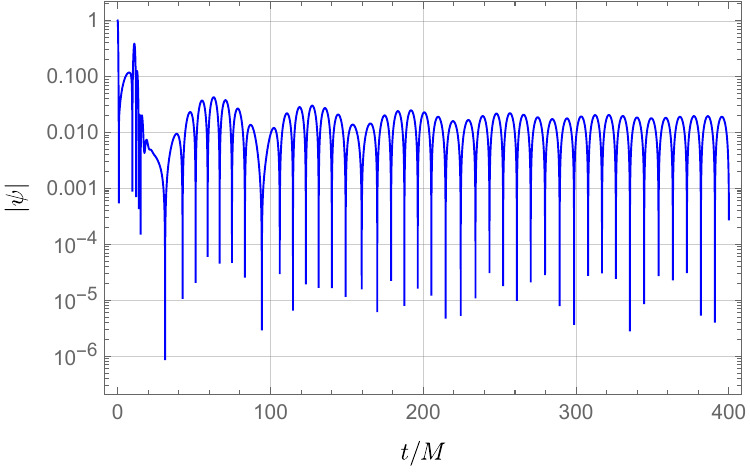}
\caption{Dynamical evolution of a Gaussian wavefunction of width $\sigma= 0.25M$ centered and observed at $v_c=r_o=10M$ for $a=0.82M$. Echoes can be observed as a modulation of the fundamental QNM  domination regime.}
\label{fig_echoes}
\end{figure}

\begin{figure}[t!]
  \includegraphics[width=0.9\columnwidth]{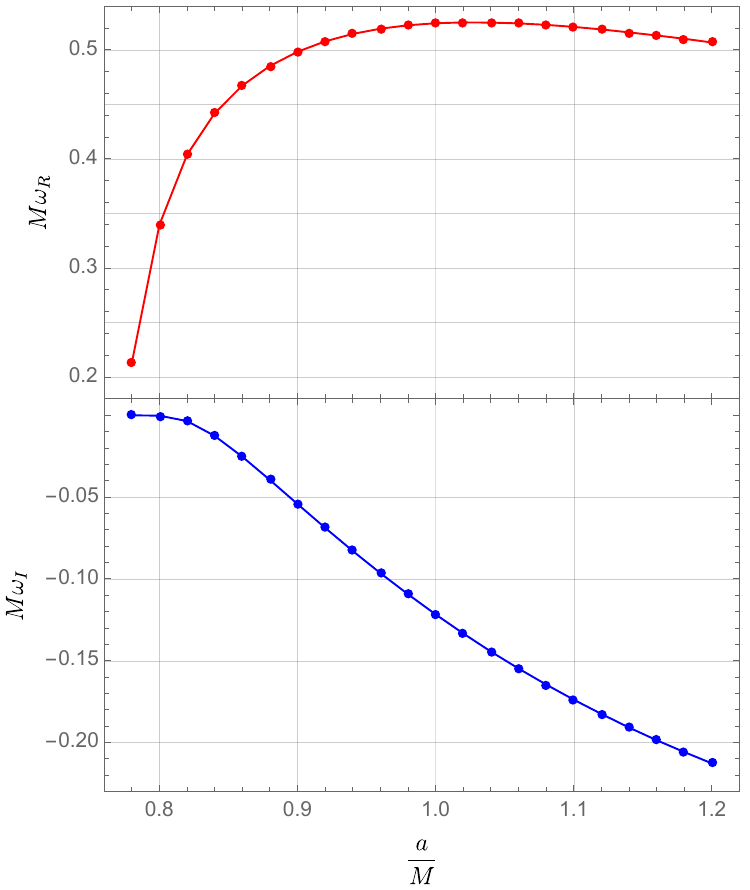}
\caption{Real part (top, red) and imaginary part (bottom, blue) of the QNMs as a function of the parameter $a$ in the absence of horizons ($a>a_\text{I}$). They have been calculated through the time-evolution of a Gaussian wave-packet of width $\sigma= 0.25M$ centered at $v_c=10M$, and observed at $r_o^*=10 M$.}
\label{fig_horizonless}
\end{figure}

\begin{figure}[t!]
  \includegraphics[width=0.9\columnwidth]{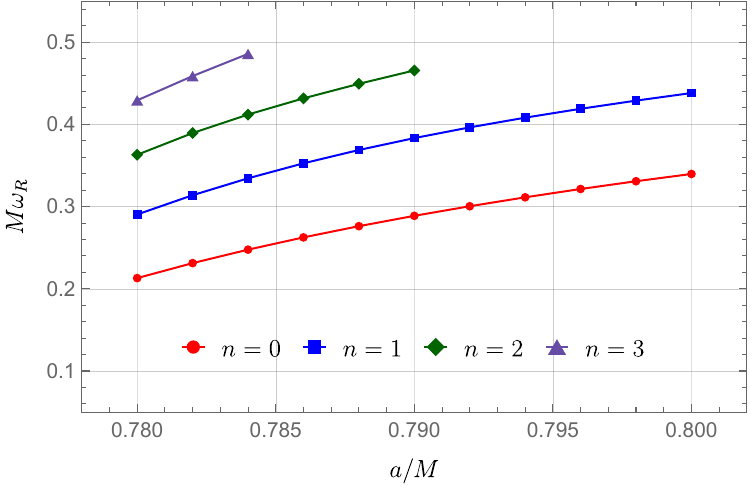}
\caption{Real part of the trapped modes (labeled by the overtone number $n$, where $n=0$ is the fundamental QNM frequency) in a range of the echo-producing cases, $a_I<a \leq 0.8$. As the potential well gets shallower, higher overtones are no longer trapped modes.}
\label{fig_overtones}
\end{figure}

When the space-time has no horizons, i.e. $a \gtrsim a_\text{I}$, we start observing {\it echoes}, i.e., secondary wave emissions. The origin of these echoes lies on the shape of the potential (see Fig. \ref{fig_peaks}), which has multiple peaks, namely, a large central peak, and two secondary and smaller symmetric peaks at each side of the throat. Such multi-peak structure creates a well on each side of the wormhole throat, allowing to trap modes. This fact  causes a  modulation of the wavefunction oscillation, as seen in Fig. \ref{fig_echoes}. The reflected wave has less amplitude, and the real part of the frequency is also diminished (see Fig. \ref{fig_horizonless}) due to the fact that higher frequencies are more likely to cross the secondary potential barrier, while lower frequencies are more likely to be reflected at it \cite{Cardoso:2016oxy}. Therefore, as time goes by, waves are reflected back and forth in the potential well and a series of echoes are observed. At late times, the modes that remain trapped in the well are those with the lowest real frequency. These modes are considered quasi-stationary, since their imaginary part is significantly small (in absolute value), and they correspond to the fundamental mode and a few overtones (see Fig. \ref{fig_overtones}), depending on the depth of the finite well. As these states are long-lived, we are able to identify them using the Prony fitting method, since they are present after the transitory phase of the time domain profile.


For values of $a$ near the extremal black hole case, $a \gtrsim a_{I}$, we observe that the secondary peaks are far from the center of the potential, but they get closer as we increase the value of $a$, as can be seen in the inset of Fig. \ref{fig_peaks}. On the other hand, as we get close to the second limiting value, $a \lesssim a_{II}$, with $a_{II}=2\sqrt{5}/5M \approx 0.894$, the local and central peaks of the potential gets very close to each another, leaving a shallower potential well for echoes to be produced. And finally, above this value we only have the central peak and echoes are no longer observed. This can be put in correspondence with the effective potential of null geodesics and, in particular, with the unstable bound orbits associated to the maxima of the potential, to be discussed in Sec. \ref{ShadowsCorresp} below.

Regarding the QNM frequencies in that regime, we observe that, near the critical value of $a=a_{I}$, the quasi-normal modes are long-lived, i.e. the imaginary part of the QNM frequency is near to zero. However, they start growing (in absolute value) as we increase $a$. The real part of the QNM  frequencies, otherwise, starts growing at first, but seems to reach a maximum near $a=1.02M$.

Let us point out that in order to obtain the QNM frequencies in the above regime we have used $h=0.01$ and $u, v \in [0, 1200M]$ for $a \in (a_I,0.84M]$, $h=0.005$ and $u, v \in [0, 300M]$ for $a \in (0.84M,a_{II}]$, and $h=0.001$ and $u, v \in [0, 120M]$ for $a \in (a_{II},1.2M]$.

\section{Correspondence of echoes with optical images}
\label{ShadowsCorresp}

The QNMs of the family of compact objects studied here can be put into correspondence with features of their optical images  via an intriguing similarity between these two seemingly different phenomena. Consider the Regge-Wheeler potential (\ref{eq:RWpot}). In the eikonal limit, $l \gg n$, the potential reduces simply to (recall that $n$ is the overtone number and $l$ the multipole number) \cite{Iyer:1986np} 
\begin{equation} \label{eq:VQNM}
V_{QNM}(r)\overset{l \gg n}{\approx } A \left(\frac{l(l+1)-2}{\Sigma^2(r)} \right) + \mathcal{O}\left(\frac{1}{\Sigma^2} \right)
\end{equation}
where terms in $\mathcal{O}(\Sigma^{-2})$ come from potential contributions of the second line of Eq.(\ref{eq:RWpot}); for instance, for the Schwarzschild solution such terms appear as $A/r^2 + \mathcal{O}(r^{-3}) $, which cancel out the term in $-2A/r^2$ in Eq.(\ref{eq:VQNM}).

This potential has the same functional form as the one followed by null trajectories (photons) in a spherically symmetric background of the form (\ref{eq:lineel}). Indeed, such equations for light propagation can be written as
\begin{equation}
AB^{-1} \left(\frac{dr}{d\lambda} \right)^2=\frac{1}{b^2}-V_{BHI}(r),
\end{equation}
where $\lambda$ is the affine parameter, $b \equiv L/E$ is the ratio between the particle's (conserved) angular momentum and energy, and known as the {\it impact parameter}, and
\begin{equation}
V_{BHI}(r)=\frac{A(r)}{\Sigma^2(r)}
\end{equation}
is the effective potential arising in black hole imaging (BHI).

In this framework, photon spheres are those defined by unstable null bound orbits, that is, orbits which asymptote, when traced backwards from the observer's screen, to the maximum of the potential. This is possible provided that their impact parameter satisfies the following equations (hereafter we drop the label BHI for simplicity)
\begin{equation}
b_c=V^{-1}(r_{ps
}) ;\ V'(r) \vert_{r=r_{ps}}=0 ;\ V''(r) \vert_{r=r_{ps}}<0.
\end{equation}
The first condition defines turning point locations, i.e., photons that are scattered back by the potential barrier, the second condition the extrema character of the potential, and the third  its maximum character. Orbits that fulfill the two first conditions but not the third (i.e. minima of the potential), are called instead as anti-photon spheres. The impact parameter itself for these photon spheres is dubbed as the critical impact parameter and can be found explicitly as
\begin{equation}
b_c=\frac{\Sigma_{ps}}{\sqrt{A_{ps}}}.
\end{equation}
where ps denotes here quantities evaluated at the photon sphere, e.g., $A_{ps} \equiv A(r_{ps})$.

\begin{figure*}[t!]
\includegraphics[width=0.8\columnwidth]{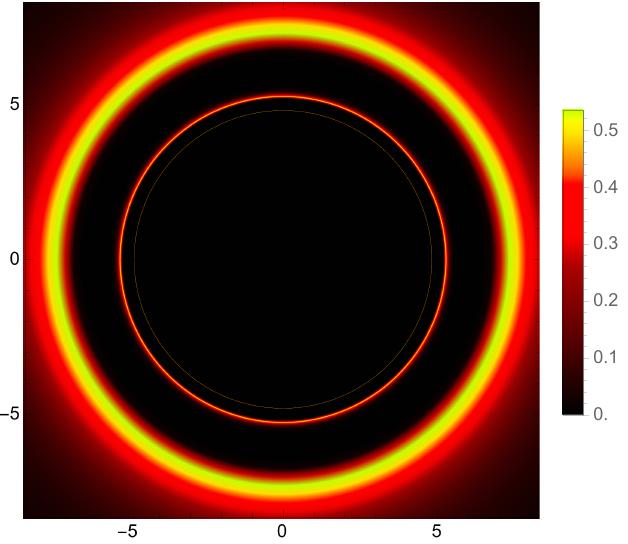}
\includegraphics[width=0.8\columnwidth]{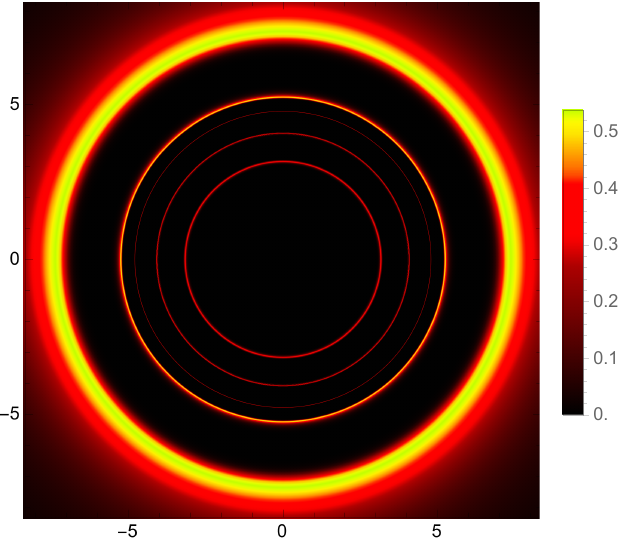}
\caption{Images of a modified black hole with $a=0.76M$ (left) and of a traversable wormhole with $a=0.78M$ (right) for the generalized black bounce proposal within a SU emission profile that peaks in the vicinity of the innermost stable circular orbit of time-like observers.}
\label{fig_horizon}
\end{figure*}

Imaging a compact object implies ray-tracing all trajectories within a certain range of parameters $b \in (0,b_{max})$, with $b_{max}>b_c$ in order to obtain a relation $r_m=r_m(b)$. The latter relation is understood as follows. Photon trajectories with $b \gtrsim b_c$ suffer heavy deflections to the point that they may wind around the photon sphere from one to many times before being released towards asymptotic infinity. Such photon trajectories produce, on the observer's screen, a series of photon rings indexed by an integer number $m=0,1,2,3, \ldots$ associated to photons that have turned a number $m$ of half turns ($m=0$ corresponding to the disk's direct emission) around the central object \cite{Gralla:2019xty}. Such rings (in the black hole case) are governed by various critical exponents \cite{Kocherlakota:2023qgo} entailing, in particular, an exponential decrease in their widths, diameters and luminosities. The relation $r_m =r_m(b)$ thus links the impact parameter of any such photon with the part(s) of the disk each one crosses it and allows to trade impact parameters by radial coordinates. Furthermore, such a set of photon rings approaches, in the limit $n \to \infty$, a critical curve, corresponding to the projection on the observer's screen of the photon sphere. The inner region to it is typically dubbed as the {\it shadow} due to the much shorter path-length of those photons intersecting the event horizon, though one must bear in mind that the actual brightness depression can be much reduced for thin and thick accretion disks \cite{Chael:2021rjo}.

Such photon rings and shadow play a key role in the correspondence between QNMs and black hole images \cite{Pedrotti:2024znu}. Such a correspondence actually equals the real and imaginary parts of the QNM with features of the imaging side following the rule
\begin{equation}
\omega_R + i \omega_I = \left(l+\frac{1}{2} \right) \Omega_c - i \left(n + \frac{1}{2} \right)  \vert \lambda \vert, 
\end{equation}
where $\Omega_c$ is the angular velocity of circular orbits while $\lambda$ sets the time-scale of null orbits traveling around the black hole \cite{Cardoso:2008bp}. The latter can be re-cast in terms of the number of half-orbits $m$ instead, providing the relation $\gamma_{ps}=\pi b_c \vert \lambda \vert$. Such quantities can be linked to observable quantities, namely, the shadow's radius (i.e. the inner region to the critical curve on the observer's screen) and the exponential fall-off (for their widths, diameters, and luminosities) of photon rings, respectively (see the discussion of  \cite{Pedrotti:2025upg}).


For the sake of our work and to illustrate the correspondence of the echoes found in the previous section with the corresponding optical images, we consider an optically and geometrically thin accretion disk featuring in the disk's own frame a monochromatic emission. The emission profile (in $r$) corresponds to a specific implementation of Johnson's Standard Unbound profiles, conceived to reproduce simulated images of the accretion flow via semi-analytical models in as a simplified setting as possible (see e.g. the analysis of  \cite{Vincent:2022fwj}) without losing a precise enough determination of the most salient features of images (the photon ring and the shadow). The particular setting employed here to generate images has been thoroughly studied e.g. in Refs. \cite{Olmo:2023lil,daSilva:2023jxa,DeMartino:2023ovj}. Here we shall focus on its GLM3 model, which peaks near the innermost stable orbits for time-like observers, and which merits in displaying the photon rings isolated from each another (as opposed to models extending up to the event horizon), facilitating their visual appearance. For the sake of comparison with our analysis below, a Schwarzschild black hole is characterized by the following quantities (in units of $M$)
\begin{equation}
r_h=2  ;\ r_{ps} =3 ;\  b_c = 3\sqrt{3}  \approx 5.196;\ \gamma_{ps} =\pi  .
\end{equation}
where we point out that the Lyapunov exponent can be computed for any general spherically symmetric metric \cite{Cardoso:2008bp}, or read off directly (but approximately) from the ray-tracing procedure.

In Fig. \ref{fig_horizon} we provide the resulting images for a generalized black bounce black hole with $a=0.76M$ and a traversable wormhole with $a=0.78M$. Differences are immediately noticeable between them. 

In the black hole case, which is characterized by the quantities (in units of $M$)
\begin{equation}
r_h \approx 1.237 ;\ r_{ps} \approx 2.443 ;\ b_c \approx 4.773 ;\ \gamma_{ps} \approx 2.488,
\end{equation}
we find the main ring of radiation (appearing in yellowish color in Fig. \ref{fig_horizon}) corresponding to the direct emission, enclosing the $m=1$ photon ring (clearly visible, appearing in reddish color) and the $m=2$ photon ring (barely visible) inner to it. Further higher-order rings would appear too demagnified to contribute to the image and, therefore, for practical purposes the $m=2$ ring here plays the role of the outer edge of the central brightness depression.  This drastic fall-off in the luminosity of successive photon rings is in agreement with the theoretical expectations based on the Lyapunov exponent, which amounts to a decrease of a factor $\propto e^{\gamma_{ps}} \approx 12.04 $  with every $m$, a very significant reduction with respect to the Schwarzschild solution, $\sim 23.14$. For the sake of simulations is therefore enough to consider up to $m=2$ light trajectories in the computation of images. However, in practice the fact that each successive light trajectory travels different regions of the disk collecting different luminosities from a non-homogeneuous disk will entail certain deviations with respect to them, to be found in a case-by-case basis. For instance, in the present modified black hole case with the GLM3 model above, the decrease of luminosity from the $m=1$ to the $m=2$ photon ring amounts to $\sim 17.98$, significantly higher than the prediction by the theoretical Lyapunov exponent.

Things are quite different in the traversable wormhole case. It is characterized by (in units of $M$)
\begin{equation}
r_t=a=0.78;\ r_{ps} \approx 2.398 ;\ b_c \approx 4.74 ;\ \gamma_{ps} \approx 2.414,  
\end{equation}
where $r_t$ is the location of the wormhole throat. In this case, due to the potential well, there are multiple additional light trajectories that may contribute to the image, associated to impact parameters  $b \gtrsim b_c$ either in their trip in or on their way back out after hitting the potential barrier. Computing trajectories up to $m=3$ in the background of the wormhole, we find two additional visible photon rings (besides those already present in the black hole) in the optical images, alongside a strongly reduced size of central brightness depression. This is due to the fact that the theoretical expectations based on the Lyapunov exponent are not met in this case. Indeed, as is obvious from its visual appearance, inner rings may be even more luminous than some of the outer ones!. Furthermore, this effect may keep producing further additional, non-negligible, rings if more higher-order trajectories $m>3$ are included in the ray-tracing procedure. The presence of non-negligible higher-order rings is consistent with the results found in the literature for several kinds of ultra-compact objects, see e.g. \cite{Olmo:2021piq}, but breaks the  expectations on their luminosity based on a direct employ of the Lyapunov exponent.

While such additional photon rings and reduced size of the central brightness depression may appear as clear observational discriminators between black hole and traversable wormhole images, one must bear in mind that the time delay experienced by those light rays exploring the innermost regions of the wormhole may drastically affect their actual luminosity in time-averaged images, as discussed in \cite{Chen:2024ibc}. Furthermore, the identification of the features of each echo with those additional photon rings seems not to be available from the current formulation of the correspondence, thus requiring further work on the subject to fully exploit the opportunities present in horizonless, ultra-compact objects for QNMs and shadows.

\section{Conclusion and discussion}
\label{conclusions}

In this work we have studied the quasi-normal modes of a family of generalized black bounces characterized by a single extra parameter $a$. These models are defined as an extension of the Schwarzschild metric which, in particular, provides a bounce in the radial function, this way preventing the focusing of geodesics which appear in the singularity theorems within GR. Furthemore, such solutions smoothly interpolate between regular black holes and traversable wormholes at a certain value of $a=a_I$, while a second value $a=a_{II}$ marks the transition at which the photon sphere of unstable bound geodesics is lost. 

After providing a general decomposition of the gravitational wave equation for a generic spherically symmetric metric (i.e. with independent temporal, radial, and angular components) and by considering the perturbation expansion of a anisotropic energy-momentum tensor, we end up in the usual Schr\"odinger-like equation which we solved using a time-domain method. This provided us the QNMs for the black hole for any value of $a$. By plotting the real and imaginary parts of such a mode against the black bounce parameter $a$ we find an increase of the real part and a decrease (in absolute value) of the imaginary part. This indicates that the wave perturbations over the generalized black bounce solution oscillate at higher frequencies while being more damped and stable than their counterparts with lower $a$ (including, in particular, the usual Schwarzschild black hole).

When the generalized black bounce parameter lies in the range $a_I<a \lesssim a_{II}$ we observe the presence of gravitational wave echoes. These are caused by the presence of trapped modes in the potential well between the central maximum and the local maximum which appear when the event horizon is lost. Such modes are progressively leaked out the potential well after each successive oscillation, having a modulated amplitude and a decreasing frequency as higher-frequencies are more likely to cross the potential barrier than the low-frequencies ones. For these echoes, as $a$ grows larger the real part of the QNM increases until a maximum is attained, while the imaginary part (in absolute value) smoothly decreases. 

We furthermore explored the correspondence recently discussed in the literature between QNMs and black hole imaging. Such a correspondence comes from the fact that, in the eikonal limit $l \gg n$ of the WKB approximation, the real and imaginary parts of the QNM can be linked to quantities appearing in black hole imaging. In fact, in such a limit the unstable surfaces giving rise to such a phenomenon lie at the same location. These facts allows to compare, for black holes, the features of QNMs (real and imaginary parts) with those of black hole imaging (shadow and photon rings). Such a comparison is somewhat blurry in the case of traversable wormholes, given the fact that echoes are recorded as time goes by, while photon rings are the product of time-averaged images. Nonetheless, using a simplified analytical setting of a thin accretion disk, we managed to find the additional photon rings that appear in the traversable wormhole images, counterparts of echoes in the imaging side.

To conclude, though the analysis carried out here makes use of a toy-model, its usefulness lies in the smooth transition it provides between regular black holes and traversable wormholes, allowing us to inspect in a simplified setting the expected changes when moving from one type of solution to another. In particular, it is uncertain how to use the QNM-BHI correspondence for horizonless compact objects to extract properties on each side, since the quantities weaved together in the correspondence not necessarily fulfill the same principles as those of the black hole solutions (illustrated here with the failed predictions based on the theoretical Lyapunov exponent). While the correspondence may be of great help for each side, there is still much work to do to clarify its possibilities and limitations (see the recent discussion in \cite{Pedrotti:2025upg}), and we hope to further report on it soon.  

\section*{Acknowledgements}

ADC is funded by a pre-doctoral contract from the Predoctoral Contracts UVa 2022 co-financed by Banco Santander. This work is supported by the Spanish National
Grants PID2020-117301GA-I00, PID2022-138607NB-I00 and CNS2024-154444, funded by MICIU/AEI/10.13039/501100011033 (``ERDF A way of making Europe" and ``PGC
Generaci\'on de Conocimiento"); the Q-CAYLE project, funded by the European Union-Next Generation UE/MICIU/Plan de Recuperacion, Transformacion y Resiliencia/Junta de Castilla y Leon (PRTRC17.11); We also acknowledge financial support of the Department of Education, Junta de Castilla y León and FEDER Funds, Ref. CLU-2023-1-05.  



\begin{thebibliography}{}

\bibitem{Addazi:2021xuf}
A.~Addazi,  \textit{et al.}
Prog. Part. Nucl. Phys. \textbf{125} (2022), 103948

\bibitem{AlvesBatista:2023wqm}
R.~Alves Batista, \textit{et al.}
Class. Quant. Grav. \textbf{42} (2025) 032001.

\bibitem{Cardoso:2019rvt}
V.~Cardoso and P.~Pani,
Living Rev. Rel. \textbf{22} (2019) 4.


\bibitem{Kerr:1963ud}
R.~P.~Kerr,
Phys. Rev. Lett. \textbf{11} (1963) 237.

\bibitem{Becerra-Vergara:2021gmx}
E.~A.~Becerra-Vergara, C.~R.~Arg\"uelles, A.~Krut, J.~A.~Rueda and R.~Ruffini,
Mon. Not. Roy. Astron. Soc. \textbf{505} (2021) L64.

\bibitem{Bambi:2016sac}
C.~Bambi, A.~Cardenas-Avendano, T.~Dauser, J.~A.~Garcia and S.~Nampalliwar,
Astrophys. J. \textbf{842} (2017) 76. 

\bibitem{Carullo:2018sfu}
G.~Carullo,  \textit{et al.}
Phys. Rev. D \textbf{98}  (2018) 104020.


\bibitem{Isi:2019aib}
M.~Isi, M.~Giesler, W.~M.~Farr, M.~A.~Scheel and S.~A.~Teukolsky,
Phys. Rev. Lett. \textbf{123} (2019) 111102.



\bibitem{LIGOScientific:2020tif}
R.~Abbott \textit{et al.} [LIGO Scientific and Virgo],
Phys. Rev. D \textbf{103} (2021) 122002. 


\bibitem{EventHorizonTelescope:2021dqv}
P.~Kocherlakota \textit{et al.} [Event Horizon Telescope],
Phys. Rev. D \textbf{103} (2021) 104047 .


\bibitem{EventHorizonTelescope:2022xqj}
K.~Akiyama \textit{et al.} [Event Horizon Telescope],
Astrophys. J. Lett. \textbf{930} (2022) L17. 

\bibitem{Senovilla:2014gza}
J.~M.~M.~Senovilla and D.~Garfinkle,
Class. Quant. Grav. \textbf{32} (2015) 124008.

\bibitem{Ayon-Beato:1998hmi}
E.~Ayon-Beato and A.~Garcia,
Phys. Rev. Lett. \textbf{80} (1998) 5056.

\bibitem{Carballo-Rubio:2019fnb}
R.~Carballo-Rubio, F.~Di Filippo, S.~Liberati and M.~Visser,
Phys. Rev. D \textbf{101} (2020) 084047.


\bibitem{Torres:2022twv}
R.~Torres,
[arXiv:2208.12713 [gr-qc]].

\bibitem{Lan:2023cvz}
C.~Lan, H.~Yang, Y.~Guo and Y.~G.~Miao,
Int. J. Theor. Phys. \textbf{62} (2023) 202. 

\bibitem{Bueno:2024dgm}
P.~Bueno, P.~A.~Cano and R.~A.~Hennigar,
Phys. Lett. B \textbf{861} (2025) 139260.

\bibitem{Alencar:2025jvl}
G.~Alencar, A.~Duran-Cabac\'es, D.~Rubiera-Garcia and D.~S\'aez-Chill\'on G\'omez,
Phys. Rev. D \textbf{111} (2025) 104020.


\bibitem{Eichhorn:2021iwq}
A.~Eichhorn and A.~Held,
JCAP \textbf{05} (2021) 073

\bibitem{Stefanov:2010xz}
I.~Z.~Stefanov, S.~S.~Yazadjiev and G.~G.~Gyulchev,
Phys. Rev. Lett. \textbf{104} (2010) 251103

\bibitem{Jusufi:2019ltj}
K.~Jusufi,
Phys. Rev. D \textbf{101} (2020) 
 084055.


\bibitem{Pedrotti:2024znu}
D.~Pedrotti and S.~Vagnozzi,
Phys. Rev. D \textbf{110} (2024) 084075.

\bibitem{Konoplya:2024lir}
R.~A.~Konoplya and A.~Zhidenko,
JCAP \textbf{09} (2024)  068.

\bibitem{Konoplya:2024vuj}
R.~A.~Konoplya and A.~Zhidenko,
Phys. Lett. B \textbf{861} (2025) 139288

\bibitem{Bisnovatyi-Kogan:2022ujt}
G.~S.~Bisnovatyi-Kogan and O.~Y.~Tsupko,
Phys. Rev. D \textbf{105} (2022) 
064040.

\bibitem{Staelens:2023jgr}
S.~Staelens, D.~R.~Mayerson, F.~Bacchini, B.~Ripperda and L.~K\"uchler,
Phys. Rev. D \textbf{107} (2023) 124026.

\bibitem{Carballo-Rubio:2024uas}
R.~Carballo-Rubio and A.~Eichhorn,
Int. J. Mod. Phys. D \textbf{33} (2024) 2441023.

\bibitem{Murk:2024nod}
S.~Murk and I.~Soranidis,
Phys. Rev. D \textbf{110} (2024) 044064.

\bibitem{Cardoso:2016oxy}
V.~Cardoso, S.~Hopper, C.~F.~B.~Macedo, C.~Palenzuela and P.~Pani,
Phys. Rev. D \textbf{94} (2016) 084031.

\bibitem{Cardoso:2017cqb}
V.~Cardoso and P.~Pani,
Nature Astron. \textbf{1} (2017)  586.

\bibitem{Cunha:2022gde}
P.~Cunha, V.P., C.~Herdeiro, E.~Radu and N.~Sanchis-Gual,
Phys. Rev. Lett. \textbf{130} (2023) 061401.

\bibitem{Marks:2025jpt}
G.~A.~Marks, S.~J.~Staelens, T.~Evstafyeva and U.~Sperhake,
[arXiv:2504.17775 [gr-qc]].

\bibitem{Simpson:2018tsi}
A.~Simpson and M.~Visser,
JCAP \textbf{02} (2019) 042 .


\bibitem{Lobo:2020ffi}
F.~S.~N.~Lobo, M.~E.~Rodrigues, M.~V.~de Sousa Silva, A.~Simpson and M.~Visser,
Phys. Rev. D \textbf{103} (2021) 084052.

\bibitem{Mazza:2021rgq}
J.~Mazza, E.~Franzin and S.~Liberati,
JCAP \textbf{04} (2021) 082.

\bibitem{Franzin:2023slm}
E.~Franzin, S.~Liberati and V.~Vellucci,
JCAP \textbf{01} (2024) 020

\bibitem{Ellis:1973yv}
H.~G.~Ellis,
J. Math. Phys. \textbf{14} (1973) 104.

\bibitem{Bronnikov:2021uta}
K.~A.~Bronnikov and R.~K.~Walia,
Phys. Rev. D \textbf{105} (2022) 044039. 



\bibitem{Maggiore:2018sht}
M.~Maggiore,
\textit{Gravitational Waves. Vol. 2: Astrophysics and Cosmology,}
Oxford University Press, 2018.

\bibitem{Regge:1957td}
T.~Regge and J.~A.~Wheeler,
Phys. Rev. \textbf{108} (1957) 1063.


\bibitem{Zerilli:1970wzz}
F.~J.~Zerilli,
Phys. Rev. D \textbf{2} (1970)  2141.


\bibitem{Sago:2002fe}
N.~Sago, H.~Nakano and M.~Sasaki,
Phys. Rev. D \textbf{67} (2003) 104017.







\bibitem{OGRe}
B.~Shoshany,
J. Open Source Softw. \textbf{6} (2021) 3416.


\bibitem{Maggiore:2007ulw}
M.~Maggiore,
\textit{``Gravitational Waves. Vol. 1: Theory and Experiments,''}
Oxford University Press, 2007.

\bibitem{Feng:2024ygo}
X.~H.~Feng and J.~Peng,
Phys. Rev. D \textbf{110} (2024) 6.

\bibitem{Konoplya:2024lch}
R.~A.~Konoplya and O.~S.~Stashko,
Phys. Rev. D \textbf{111} (2025) 104055.

\bibitem{Berti:2009kk}
E.~Berti, V.~Cardoso and A.~O.~Starinets,
Class. Quant. Grav. \textbf{26} (2009) 163001.


\bibitem{Konoplya:2011qq}
R.~A.~Konoplya and A.~Zhidenko,
Rev. Mod. Phys. \textbf{83} (2011) 793.

\bibitem{Pani:2013pma}
P.~Pani,
Int. J. Mod. Phys. A \textbf{28} (2013) 1340018.


\bibitem{Franchini:2023eda}
N.~Franchini and S.~H.~V\"olkel,
[arXiv:2305.01696 [gr-qc]].

\bibitem{DuttaRoy:2019zvw}
P.~Dutta Roy, J.~Das and S.~Kar,
Eur. Phys. J. Plus \textbf{134} (2019) 571.


\bibitem{Gundlach:1993tp}
C.~Gundlach, R.~H.~Price and J.~Pullin,
Phys. Rev. D \textbf{49} (1994) 883.



\bibitem{Molina:2010fb}
C.~Molina, P.~Pani, V.~Cardoso and L.~Gualtieri,
Phys. Rev. D \textbf{81} (2010) 124021.

\bibitem{Chavda:2024awq}
A.~Chavda, M.~Lagos and L.~Hui,
[arXiv:2412.03435 [gr-qc]].


\bibitem{Konoplya:2023ahd}
R.~A.~Konoplya, D.~Ovchinnikov and B.~Ahmedov,
Phys. Rev. D \textbf{108} (2023) 104054.





\bibitem{Kyutoku:2022gbr}
K.~Kyutoku, H.~Motohashi and T.~Tanaka,
Phys. Rev. D \textbf{107} (2023) 044012.


\bibitem{Nollert:1999ji}
H.~P.~Nollert,
Class. Quant. Grav. \textbf{16} (1999) R159.



\bibitem{Leaver:1986gd}
E.~W.~Leaver,
Phys. Rev. D \textbf{34} (1986) 384.



\bibitem{Berti:2007dg}
E.~Berti, V.~Cardoso, J.~A.~Gonzalez and U.~Sperhake,
Phys. Rev. D \textbf{75} (2007), 124017













\bibitem{Mamani:2022akq}
L.~A.~H.~Mamani, A.~D.~D.~Masa, L.~T.~Sanches and V.~T.~Zanchin,
Eur. Phys. J. C \textbf{82} (2022)  897.


\bibitem{Chandrasekhar:1975zza}
S.~Chandrasekhar and S.~L.~Detweiler,
Proc. Roy. Soc. Lond. A \textbf{344} (1975) 441.


\bibitem{Iyer:1986np}
S.~Iyer and C.~M.~Will,
Phys. Rev. D \textbf{35}, 3621 (1987)

\bibitem{Gralla:2019xty}
S.~E.~Gralla, D.~E.~Holz and R.~M.~Wald,
Phys. Rev. D \textbf{100} (2019) 024018.

\bibitem{Kocherlakota:2023qgo}
P.~Kocherlakota, L.~Rezzolla, R.~Roy and M.~Wielgus,
Phys. Rev. D \textbf{109} (2024) 064064.

\bibitem{Chael:2021rjo}
A.~Chael, M.~D.~Johnson and A.~Lupsasca,
Astrophys. J. \textbf{918} (2021) 6.



\bibitem{Cardoso:2008bp}
V.~Cardoso, A.~S.~Miranda, E.~Berti, H.~Witek and V.~T.~Zanchin,
Phys. Rev. D \textbf{79} (2009) 064016.

\bibitem{Pedrotti:2025upg}
D.~Pedrotti and M.~Calz\`a,
[arXiv:2504.01909 [gr-qc]].

\bibitem{Vincent:2022fwj}
F.~H.~Vincent, S.~E.~Gralla, A.~Lupsasca and M.~Wielgus,
Astron. Astrophys. \textbf{667} (2022) A170.


\bibitem{Olmo:2023lil}
G.~J.~Olmo, J.~L.~Rosa, D.~Rubiera-Garcia and D.~Saez-Chillon Gomez,
Class. Quant. Grav. \textbf{40} (2023) 174002.

\bibitem{daSilva:2023jxa}
L.~F.~D.~da Silva, F.~S.~N.~Lobo, G.~J.~Olmo and D.~Rubiera-Garcia,
Phys. Rev. D \textbf{108} (2023) 084055.

\bibitem{DeMartino:2023ovj}
I.~De Martino, R.~Della Monica and D.~Rubiera-Garcia,
Phys. Rev. D \textbf{108} (2023)  124054.

\bibitem{Olmo:2021piq}
G.~J.~Olmo, D.~Rubiera-Garcia and D.~S.~C.~G\'omez,
Phys. Lett. B \textbf{829} (2022) 137045.

\bibitem{Chen:2024ibc}
C.~Y.~Chen and Y.~Yokokura,
Phys. Rev. D \textbf{109} (2024) 104058.





\end{thebibliography}
\end{document}